\documentclass[aps,prb,onecolumn,superscriptaddress,floats]{revtex4}
\usepackage{txfonts}
\usepackage{amssymb}
\usepackage{graphicx}
\usepackage{epstopdf}
\usepackage{multirow}
\usepackage{float}
\renewcommand\figurename{\textbf{Figure}}

\begin{document}
\large
\title{Structural and Magnetic Phase Diagram of CrAs and its Relationship with Pressure-induced Superconductivity
}
\author{Yao Shen$^\sharp$}
\affiliation{
State Key Laboratory of Surface Physics and Department of Physics, Fudan University, Shanghai 200433, China
}
\author{Qisi Wang$^\sharp$}
\affiliation{
State Key Laboratory of Surface Physics and Department of Physics, Fudan University, Shanghai 200433, China
}
\author{Yiqing Hao$^\sharp$}
\affiliation{
State Key Laboratory of Surface Physics and Department of Physics, Fudan University, Shanghai 200433, China
}
\author{Bingying Pan}
\affiliation{
State Key Laboratory of Surface Physics and Department of Physics, Fudan University, Shanghai 200433, China
}
\author{Yu Feng}
\affiliation{
State Key Laboratory of Surface Physics and Department of Physics, Fudan University, Shanghai 200433, China
}

\author{Q. Huang}
\affiliation{
NIST Center for Neutron Research, National Institute of Standards and Technology, Gaithersburg, Maryland 20899, USA
}
\author{L. W. Harriger}
\affiliation{
NIST Center for Neutron Research, National Institute of Standards and Technology, Gaithersburg, Maryland 20899, USA
}
\author{J. B. Leao}
\affiliation{
NIST Center for Neutron Research, National Institute of Standards and Technology, Gaithersburg, Maryland 20899, USA
}
\author{Y. Zhao}
\affiliation{
NIST Center for Neutron Research, National Institute of Standards and Technology, Gaithersburg, Maryland 20899, USA
}
\affiliation{
Department of Materials Science and Engineering, University of Maryland, College Park, Maryland 20742, USA
}
\author{R. M. Chisnell}
\affiliation{
NIST Center for Neutron Research, National Institute of Standards and Technology, Gaithersburg, Maryland 20899, USA
}
\author{J. W. Lynn}
\affiliation{
NIST Center for Neutron Research, National Institute of Standards and Technology, Gaithersburg, Maryland 20899, USA
}
\author{Huibo Cao}
\affiliation{
Neutron Scattering Science Division, Oak Ridge National Laboratory, Oak Ridge, Tennessee 37831-6393, USA
}
\author{Jiangping Hu}
\affiliation{
Institute of Physics, Chinese Academy of Sciences, Beijing 100190, China
}
\affiliation{
Department of Physics, Purdue University, West Lafayette, Indiana 47907, USA
}
\author{Jun Zhao$^\ast$}
\affiliation{
State Key Laboratory of Surface Physics and Department of Physics, Fudan University, Shanghai 200433, China
}
\affiliation{
Collaborative Innovation Center of Advanced Microstructures, Fudan University, Shanghai 200433, China
}

\maketitle
\textbf{Most unconventional superconductors, including cuprates and iron-based superconductors, are derived from chemical doping or application of pressure on their collinearly magnetic-ordered parent compounds \cite{lee,dai,stewart2,powell,scalapino}. The recently discovered pressure-induced superconductor CrAs, as a rare example of a non-collinear helimagnetic superconductor, has therefore generated great interest in understanding microscopic magnetic properties and their interplay with superconductivity \cite{wu,kotegawa1,norman3}.
Unlike cuprates and iron based superconductors where the magnetic moment direction barely changes upon doping, here we show that CrAs exhibits a spin reorientation from the $ab$ plane to the $ac$ plane, along with an abrupt drop of the magnetic propagation vector at a critical pressure ($P_c\approx0.6$ GPa). This magnetic phase transition coincides with the emergence of bulk superconductivity, indicating a direct connection between magnetism and superconductivity. With further increasing pressure, the magnetic order completely disappears near the optimal $T_c$ regime ($P\approx0.94$ GPa). Moreover, the Cr magnetic moments between nearest neighbors tend to be aligned antiparallel with increasing pressure toward the optimal superconductivity regime. Our findings suggest that the non-collinear helimagnetic order is strongly coupled to structural and electronic degrees of freedom, and that antiferromagnetic correlations associated with the low magnetic vector phase are crucial for superconductivity.}

 CrAs has a MnP-type orthorhombic crystal structure at room temperature, which can be considered as a distorted hexagonal NiAs-type structure \cite{boller,selte,watanabe,kazama,zavadskii}.
 Previous measurements have shown that the system exhibits a first order non-collinear helimagnetic phase transition accompanied by a magnetostriction below $T_N$$\approx$$270$ K \cite{boller,selte,watanabe,kazama,zavadskii}.  The resistivity of CrAs also displays a clear anomaly around the $\mathrm{N\acute{e}el}$ temperature \cite{wu,wu2,kotegawa1}. This anomaly is suppressed progressively under pressure and fades away at $P>0.6$ GPa, before bulk superconductivity appears \cite{wu,kotegawa1}. The maximum $T_c\approx2$ K is attained at $P\approx1$ GPa, above which $T_c$ decreases with increasing pressure \cite{wu,kotegawa1}. It has also been revealed that the nuclear spin-lattice relaxation rate in CrAs shows substantial magnetic fluctuations, but does not display a coherence peak in the superconducting state, indicating an unconventional pairing mechanism \cite{kotegawa2}. Although these results suggest that CrAs is the first Cr-based unconventional superconductor, it remains unclear how the lattice and non-collinear magnetic structures evolve under pressure as the system is tuned from a metallic helimagnet to a superconductor. Such information is important to establish whether and what kind of magnetism is directly associated with superconductivity in this system.

 We used neutron scattering to study the structural and magnetic phase transitions of CrAs. Fig. 1c illustrates the diffraction pattern at $5$ K at ambient pressure, which can be described by the orthorhombic $Pnma$ space group together with a non-collinear double helimagnetic structure (Fig. 1a). Similar to earlier work \cite{selte}, the Cr magnetic moment [$1.724(9)\mu_B$] lies in the $ab$ plane, and the helical propagation vector is $\vec{k}=(0, 0, 0.35643)$ along the $c$ axis at $T= 4$ K. Our magnetic refinements also suggest that the phase angle between the helixes running through $S_1$ and $S_2$ is $\beta_{12}$=$-108.8(5)^\circ$, and that the angle between $S_2$ and $S_3$ is $\beta_{23}=172.9(5)^\circ$ (Fig. 1a).

To determine the evolution of the magnetic structure as a function of pressure, we carried out neutron powder diffraction measurements in an aluminium alloy or steel pressure cell (see \textbf{Methods} for details). Fig. 1d shows the diffraction pattern in the aluminium alloy pressure cell at $P=0.4$ GPa and $T=4$ K. Magnetic refinements confirm that the magnetic structure is similar to that at ambient pressure, except for the slightly reduced propagation vector $(0, 0, 0.3171)$ and moment [$1.71(2)\mu_B$]. The most striking discovery is that the diffraction pattern at $P=0.6$ GPa (Fig. 1e) shows substantial changes of the magnetic reflections compared with those at ambient and low pressures, indicative of a magnetic phase transition. Our refinement analysis reveals that these changes are due to a spin reorientation from the $ab$ plane to the $ac$ plane together with a rapid decrease of magnetic propagation vector to $\vec{k}=(0, 0, 0.2080)$, as shown in Fig. 1b. The strongly pressure dependent propagation vector and spin rotation plane implies that the helimagnetic order mainly arises from frustrated magnetic exchange interactions between local moments. Indeed, the decrease of the magnetic propagation vector can be qualitatively understood by considering a minimum magnetic exchange model based on nearest neighbor spins and the Dzyaloshinskii-Moriya interactions. There are four spins in one unit cell, as labeled in Figs. 1a and 1b. The interaction between spin $\hat S_i$ and $\hat S_j $ can be generally written as: \begin{eqnarray}
\hat H_{ij}=  \vec D_{ij} \cdot (\hat S_i\times \hat S_j)+ J_{ij} \hat S_i\cdot \hat S_j,
\label{h12}
\end{eqnarray}
where the first term is the Dzyaloshinskii-Moriya interaction. As the component of $\vec D$ along the $c$ axis is larger than the one along the $b$ axis, the propagation vector decreases when the helix rotation plane changes from the $ab$ plane to the $ac$ plane (detailed calculations are described in the Supplementary Information). In addition, the crystallographic structure refinements suggest that the nuclear peaks can be described by an orthorhombic $Pnma$ space group for all pressures and temperatures measured. The refined structural and magnetic parameters at representative temperature and pressure are given in Table I and Table II.

 Figure 2 summarizes the structural and magnetic phase transitions under various pressures. At ambient pressure, the $(0, 0, 0)\pm$ magnetic peak abruptly disappears on warming to $T_N=272$ K, suggesting a first order magnetic phase transition (Figs. 2a and 2k). Similar behavior is also observed at $P=0.4$ GPa (Figs. 2c and 2l). At the critical pressure $P_c=0.6$ GPa, the magnetic peak shows a spin reorientation behavior from the $ab$ plane to the $ac$ plane below $T_r= 88$ K (Figs. 2e and 2m). With further increasing pressure to $P=0.72,0.82$ GPa, the magnetic moment stays in the $ac$ plane, while its magnitude is further reduced. Accompanied by the magnetic phase transitions, structural phase transitions are also observed under various pressures (Figs. 2b, 2d, 2f, 2h, 2j, and 2k-2o), even though the crystal symmetry remains unchanged below and above the $\mathrm{N\acute{e}el}$ temperature. The temperature-dependent profile of nuclear peaks shows a sudden shift near the $\mathrm{N\acute{e}el}$ temperature, indicative of discontinuous changes of lattice parameters (Figs. 2b, 2d, 2f, 2h, and 2j). The structural phase transition occurs at the same temperature as the onset of the magnetic order at all pressures measured (Figs. 2k-2o). Interestingly, we notice that the magnetic peak positions also shift progressively on cooling, indicating that the propagation vector is also temperature-dependent (Figs. 2a, 2c, 2e, 2g, and 2i). Fig. 3a summarizes the pressure and temperature dependence of the propagation vector. Obviously, the propagation vector decreases gradually with decreasing temperature at $P=0$ and $0.4$ GPa, when the magnetic moment is in the $ab$ plane (Fig. 3a), which is similar to previous results at ambient pressure \cite{selte}. On the other hand, the propagation vector actually increases slightly with decreasing temperature at $P=0.72$, $0.82$ GPa, when the magnetic moment has rotated to the $ac$ plane. At the critical pressure $P_c\approx0.6$ GPa, the propagation vector decreases drastically on cooling to below $T_r= 88$ K, owing to the spin reorientation from the $ab$ plane to the $ac$ plane (Fig. 3a). The dramatically pressure- and temperature-dependent magnetic propagation vector observed here is different from the weakly pressure- and temperature-dependent spin density wave vector of metallic chromium \cite{fawcett,feng}. Moreover, the nearest neighbour Cr-Cr bond distances ($d=2.856$, $3.090$, and $3.588$ $\AA$; $P=0$ GPa, $T=5$ K) in CrAs are larger than that ($\sim 2.498\AA $) of chromium. These results together with the fact that the magnetic moment of CrAs is much larger than in metallic chromium ($\sim0.6\mu_B$, ref. \onlinecite{fawcett}) suggest that Cr moments are more localized in CrAs.

  To summarize the data in Figs. 1, 2 and 3a, we plot in Fig. 3b the structural and magnetic phase diagram of CrAs under pressure along with the superconducting transition temperatures determined from susceptibility measurements \cite{wu}. Both the magnetic and structural phase transition temperature and the magnetic moment are gradually suppressed by pressure and eventually completely disappear at $P\approx0.94$ GPa, where optimal $T_c$ ($T_c$ is maximal) is realized (Figs. 3b and 3c). In contrast to previous resistivity measurements that suggest that the magnetic order is completely suppressed above $P>0.6$ GPa \cite{wu,kotegawa1,zavadskii}, our neutron data show that the magnetic order and structural distortion persist at $P=0.72$, $0.82$, and $0.88$ GPa, where the resistivity anomaly disappears (Fig. 3b). As the pressure dependence of the propagation vector at $4$ K shows a sudden drop at $P_c=0.6$ GPa (Fig. 3d), which is accompanied by the spin reorientation, the diminishing resistivity anomaly is very likely due to the spin reorientation transition. Interestingly, an anomaly on the $\mathrm{N\acute{e}el}$ temperature vs. pressure curve is also observed at $P=0.72$ GPa (Fig. 3b), probably due to the competition between the high propagation vector phase and the low propagation vector phase near the spin reorientation transition. More importantly, as shown in Fig. 3b, the contour propagation vector map and the superconducting transition temperature plot reveal a close connection between superconductivity and the low propagation vector phase.
  Comparison of the propagation vector with the superconducting volume fraction as a function of pressure further proves that the emergence of bulk superconductivity is directly associated with the spin reorientation and the subsequent decrease of the propagation vector (Fig. 3d). These results indicate strong coupling between magnetism and superconductivity in this system.

  Figure 4 illustrates the pressure effect of the lattice constants, bond distances, and angles between Cr moments at various temperatures obtained from our refinement analysis. At ambient pressure, a large magnetostriction ($\triangle b/b$=$3.45\%$, $\triangle a/a$=$-0.29\%$, $\triangle c/c$=$-0.84\%$, $\triangle v/v$=$2.28\%$, $P=0$ GPa) is observed below the $\mathrm{N\acute{e}el}$ temperature (Figs. 4a-4c). The increased volume of the unit cell below $T_N$ naturally suggests that the magnetic order could be suppressed by reducing magnetostriction under pressure.
  However, as shown in Figs. 4a-4c, the magnetostriction ($\triangle b/b$=$5.69\%$, $\triangle a/a$=$-0.88\%$, $\triangle c/c$=$-1.36\%$, $\triangle v/v$=$3.33\%$, $P=0.6$ GPa) actually increases with increasing pressure. On the other hand, our detailed refinement analysis suggests that pressure significantly reduces the nearest neighbor bond lengths of Cr atoms (Figs. 4e and 4f), which could make the $d$ electrons of Cr more itinerant, and therefore reduce the magnetic moment. More interestingly, the pressure dependence of the lattice parameters in the magnetically order state displays a clear anomaly at the critical pressure $P_c=0.6$ GPa (Fig. 4d), which corresponds to the spin reorientation transition and the emergence of bulk superconductivity (Figs. 3b and 3d). These observations indicate a strong coupling between electronic, magnetic and lattice degrees of freedom in this system. We also notice that the angle $\beta_{12}$ between Cr moments $S_1$ and $S_2$ decreases drastically from $\sim-100^\circ$ (almost perpendicular) to $\sim-160^\circ$ (almost antiparallel) with increasing pressure $P>0.6$ GPa (Fig. 4g). The angle $\beta_{23}$ between $S_2$ and $S_3$ is close to $180^\circ$ and barely changes with pressure (Fig. 4h).  These results indicate that the moments between nearest neighbors tend to be antiferromagnetically aligned in the bulk superconductivity regime. If superconductivity is indeed mediated by fluctuations of the magnetic order in this system, these results may suggest that antiferromagnetic fluctuations are important to superconductivity.

  It is suggestive to compare the electronic phase diagram of CrAs with that of other magnetic unconventional superconductors  \cite{lee,dai,stewart2,scalapino,norman2}. Unlike cuprates and iron based superconductors where chemical doping or application of pressure only reduces the magnitude of the magnetic moment while leaving the moment direction essentially unchanged \cite{lee,dai}, pressure induces a spin reorientation transition in CrAs, indicating a rather strong coupling between magnetic and lattice degrees of freedom. The highly tunable non-collinear magnetic structure under pressure and the large magnetostriction observed in CrAs are likely due to the frustrated magnetic interactions, which resembles the behavior of multiferroic materials with non-collinear magnetic structures \cite{iucrj}. The direct connection between the emergence of superconductivity and the spin reorientation has been observed in a ferromagnetic superconductor URhGe under external magnetic field \cite{levy}, but has not been observed in other magnetic superconductors without a net macroscopic magnetic moment. In the case of URhGe, the collinear ferromagnetic moment exhibits a reorientation towards the external magnetic field direction when superconductivity appears. The emergence of superconductivity with spin rotation in URhGe is believed to be driven by critical magnetic fluctuations and a Fermi surface Lifshitz transition near a quantum critical point induced by the external magnetic field \cite{levy,yelland}. Interestingly, the normal state resistivity of CrAs under pressure ($0.3$ GPa$\le$ $P$ $\le2.1$ GPa) follows the power-law relationship: $\rho_0+AT^{1.5}$, indicating the presence of magnetic quantum critical points \cite{wu}. It is currently unclear whether the spin reorientation observed in CrAs is associated with any Fermi surface change. However, in contrast to URhGe, the spin orientation in CrAs is also accompanied with an abrupt reduction of the magnetic propagation vector. Theoretically, it has been shown that the spin fluctuations at a relatively large wave vector tend to lead a singlet pairing while that at small wave vector close to $q=0$ would favor a triplet pairing \cite{lee,dai,stewart2,powell,scalapino,mazin,norman2,sro}. If these theories apply in CrAs, it would imply that the pairing state in CrAs could be wave vector/pressure dependent.  It is of great theoretical importance if the pairing symmetry could be tuned in a controlled manner in a helimagnetic superconductor \cite{norman3}.

   In summary, we have determined the structural and magnetic phase diagram of CrAs as the system is changed from a non-collinear helimagnetic metal to a superconductor under pressure.  We show that CrAs exhibits a spin reorientation from the $ab$ plane to the $ac$ plane, together with a rapid decrease of the magnetic propagation vector from (0, 0, 0.3564) at ambient pressure to (0, 0, 0.144) above the critical pressure ($P_c\approx0.6$ GPa), where bulk superconductivity begins to emerge. Moreover, pressure reduces the nearest-neighbor bond lengths of Cr atoms, which could make the electrons more itinerant and therefore reduce the magnetic moment. Furthermore, the nearest neighbor spins tend to be aligned antiparallel near the optimal superconductivity regime under pressure, suggesting that antiferromagnetic correlations between nearest neighbors may be essential for superconductivity. The highly tunable magnetic moment direction and  propagation vector observed in CrAs open up new avenues of research into the interplay between non-collinear helimagnetism and unconventional superconductivity.

\textbf{Methods}

 Our polycrystalline sample was synthesized as reported in ref.\onlinecite{zavadskii}. Consistent with previous measurements, the susceptibility of our sample displays a clear anomaly near $270$ K (refer to supplementary Figure S1), indicating a magnetic phase transition \cite{wu,kotegawa1}. Our neutron scattering measurements were carried out on the BT-1 powder diffractometer, SPINS cold triple-axis spectrometer, and BT-7 thermal triple axis spectrometer at the NIST Center for Neutron Research. The neutron wavelengths employed were $1.5389$ or $2.0785$ {\AA} using the Ge(311) monochromator at BT-1, $4.0449$ {\AA} using PG (002) monochromator at SPINS, and $2.36$ {\AA} using PG (002) monochromator at BT-7 (ref. \onlinecite{jeff}). The neutron-diffraction data Rietveld refinements are based on the program FULLPROF \cite{fullprof}. For the pressure effect measurements, the polycrystalline sample was loaded into an aluminium alloy or steel pressure cell, which was connected to an external piston-driven pressure intensifier via a heated capillary line, and pressurized using helium as a pressure medium. All pressure changes were executed while maintaining the pressure vessels above the PxT line of helium. For the measurements below the PxT line of helium, hydrostatic conditions were maintained by employing a ``BURP'' process (a careful pressurization procedure in order to ensure pressure homogeneity within the pressure cells).

 \textit{Note added:} After we finished this study, we became aware of a related preprint\cite{keller}, in which neutron diffraction measurements were performed at $P\le0.65$ GPa using a clamp pressure cell, and the data are similar to ours at ambient and low pressures before the spin reorientation transition.

$^{\sharp}$These authors contributed equally to this work.
$^{*}$Correspondence and requests for materials should be addressed to J.Z. (zhaoj@fudan.edu.cn.).

{\bf Acknowledgements}

This work is supported by the National Natural Science Foundation of China (Grant No. 11374059) and the Shanghai Pujiang Scholar Program (Grant No.13PJ1401100). H.C. received support from the Scientific User Facilities Division, Office of Basic Energy Sciences, US Department of Energy.


\begin{figure}[h]
\includegraphics[width=7cm]{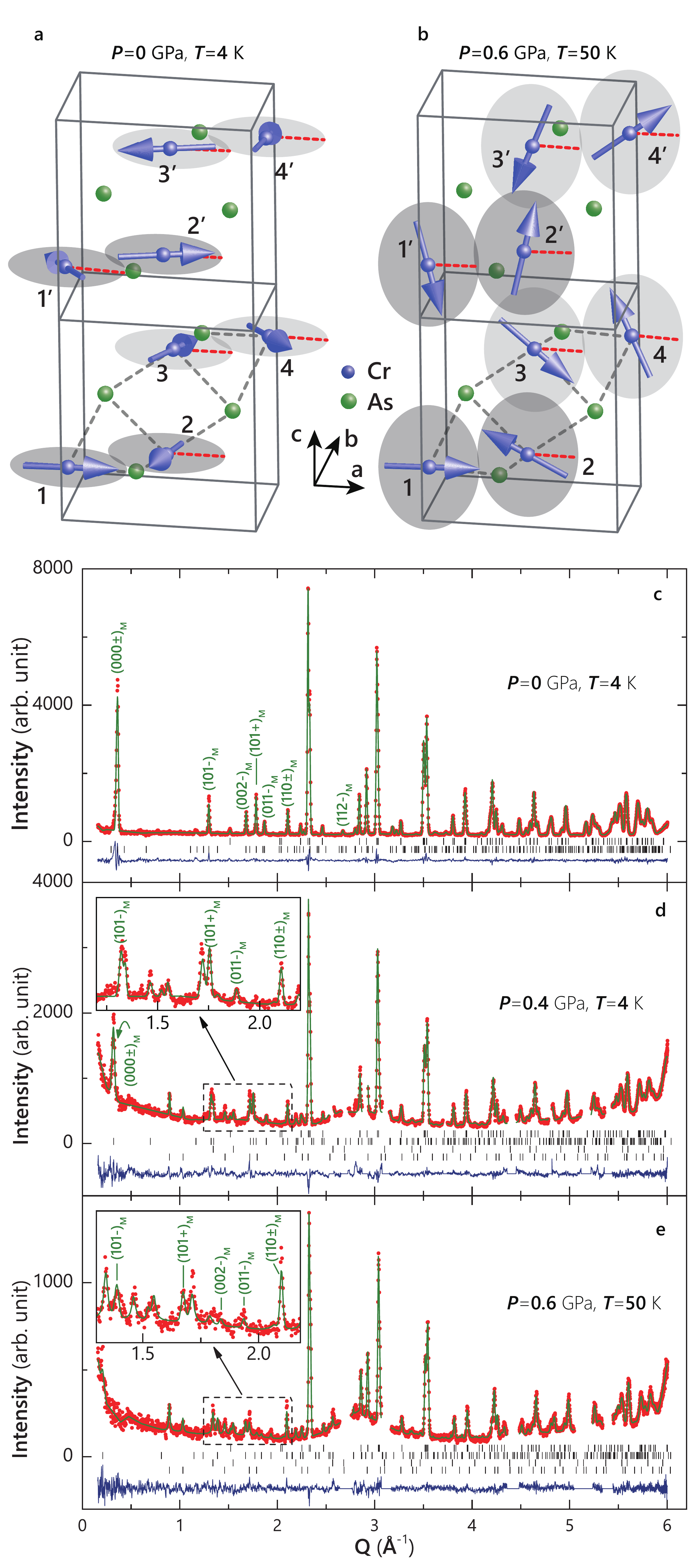}
\caption{ \textbf{Pressure dependence of the magnetic structure for CrAs}. The neutron diffraction experiments were performed on the BT-1 diffractometer using a Ge (3,1,1) monochromator. The wavelength of the incident beam was $\lambda=2.0785 {\AA}$. \textbf{a}, Magnetic structure in CrAs at ambient pressure and $T=4$ K determined by our refinements. There are four Cr atoms in one unit cell. The magnetic moments on Cr lie in the $ab$ plane. \textbf{b}, Magnetic structure of CrAs at $P=0.6$ GPa and $T=50$ K. The Cr moment is in the $ac$ plane. \textbf{c}, Observed (red) and calculated (green) neutron powder diffraction intensities of CrAs at ambient pressure and $4$ K. The spectrum is refined in the orthorhombic $Pnma$ space group together with the magnetic structure shown in Fig. 1a. Short black vertical lines show the Bragg peak positions. The purple line indicates the difference between the observed and calculated intensities. Magnetic peaks are indexed in the figure. \textbf{d}, The diffraction spectrum measured in an aluminium alloy pressure cell at $P=0.4$ GPa and $T=4$ K. Magnetic peaks are highlighted and indexed in the inset. The four refined phases are crystal and magnetic structures of CrAs, the second and third order neutron reflections from the aluminium pressure cell. The missing data correspond to the strong first order neutron reflections from the aluminium pressure cell. \textbf{e}, The diffraction spectrum measured at $P=0.6$ GPa and $T=50$ K. The spectrum is refined with the magnetic structure shown in Fig. 1b.
}
\end{figure}

 \begin{figure*}[h]
\includegraphics[width=18cm]{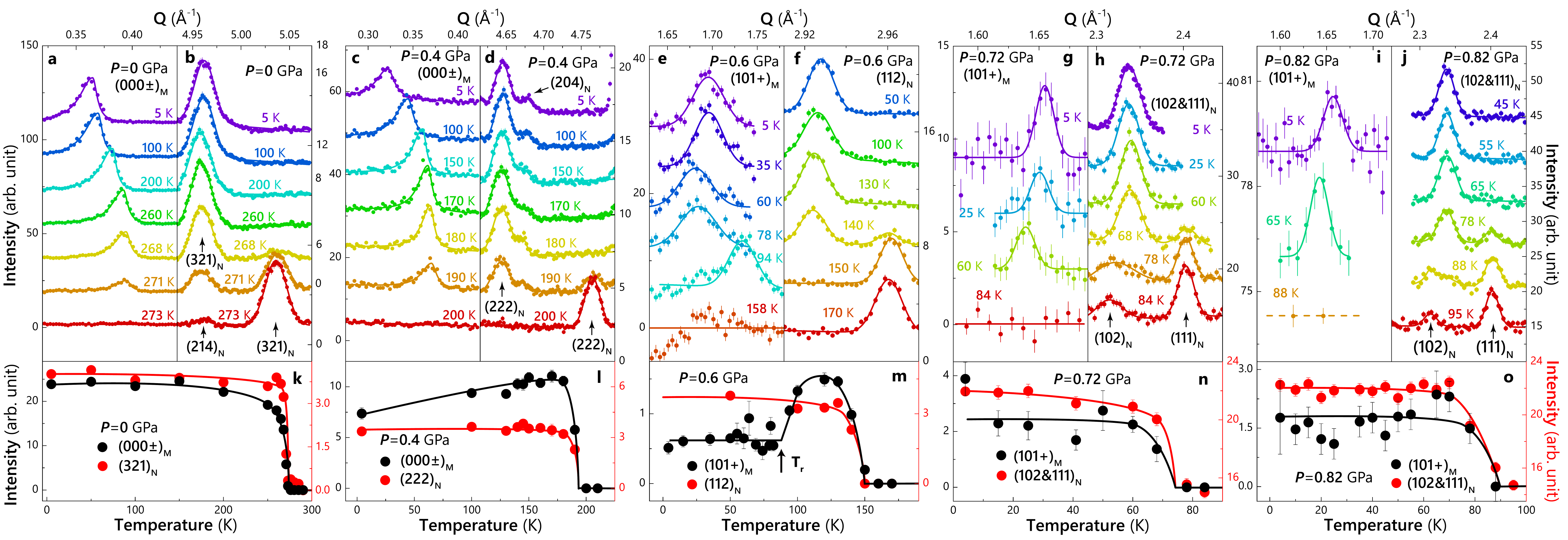}
\caption{\textbf{Structural and magnetic phase transition temperatures under various pressures for CrAs}. The data in \textbf{a-f} and \textbf{g-j} were collected in the aluminium alloy and steel pressure cell, respectively.  \textbf{a-j}, Temperature-dependent diffraction profile of the magnetic and nuclear peaks at various pressures, $(000\pm)$ magnetic peak at $P=0$ GPa, \textbf{a};  $(321)$ nuclear peak at $P=0$ GPa, \textbf{b}; $(000\pm)$ magnetic peak at $P=0.4$ GPa, \textbf{c}; $(222)$ and $(204)$ nuclear peaks at $P=0.4$ GPa, \textbf{d}; $(101+)$ magnetic peak at $P=0.6$ GPa, \textbf{e}; $(112)$ nuclear peak at $P=0.6$ GPa, \textbf{f}; $(101+)$ magnetic peak at $P=0.72$ GPa, \textbf{g};  $(102)$ and $(111)$ nuclear peaks at $P=0.72$ GPa, \textbf{h}; $(101+)$ magnetic peak at $P=0.82$ GPa, \textbf{i}; $(102)$ and $(111)$ nuclear peaks at $P=0.82$ GPa, \textbf{j}. \textbf{k-o}, temperature dependence of the peak intensity of the data obtained in \textbf{a-j}. We note that in \textbf{l}, the decrease of the $(000\pm)_M$ peak intensity at low temperature is due to the decrease of the phase angle $\beta_{12}$ on cooling, as shown in Fig. 4g. The error bars indicate one standard deviation.
 }
\label{cha}
\end{figure*}

\begin{figure}[h]
\includegraphics[width=0.7\textwidth]{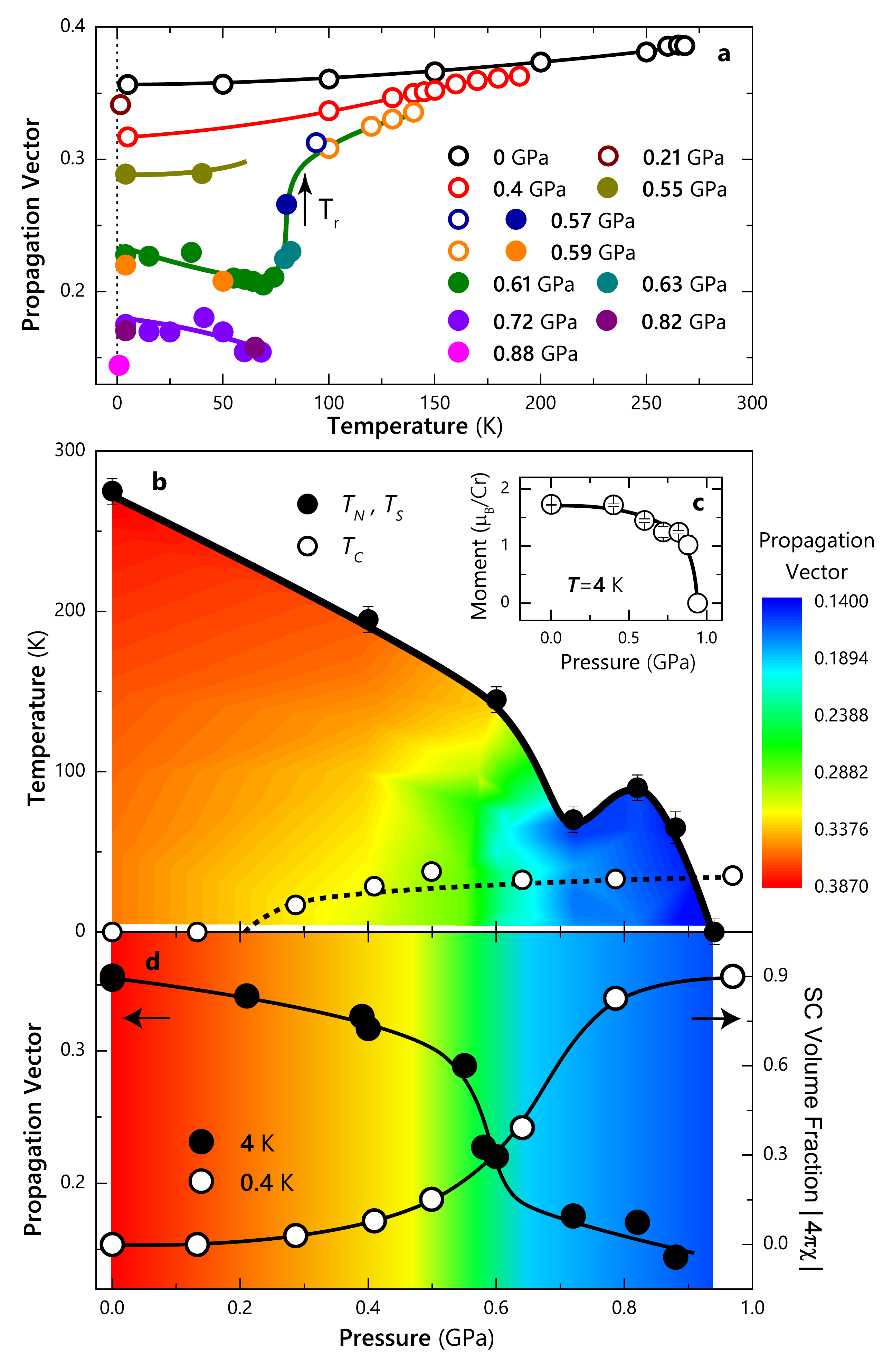}
\caption{\textbf{Temperature and pressure dependence of the magnetic propagation vector and the structural and magnetic phase diagram of CrAs.} \textbf{a}, Temperature and pressure dependence of the magnetic propagation vector. The open and filled circles indicate the magnetic moment in the $ab$ and $ac$ plane, respectively. \textbf{b}, Structural and magnetic phase diagram of CrAs. The contour propagation vector map is plotted from the data in Fig. 3a in the temperature-pressure space. The superconducting transition temperatures are adapted from ref. \onlinecite{wu}. \textbf{c}, The pressure dependence of the magnetic moment at $4$ K. \textbf{d}, The pressure dependence of the propagation vector and the superconducting volume fraction determined by susceptibility measurements adapted from ref. \onlinecite{wu}. The error bars indicate one standard deviation.
}
\end{figure}

\begin{figure}[h]
\includegraphics[scale=0.25]{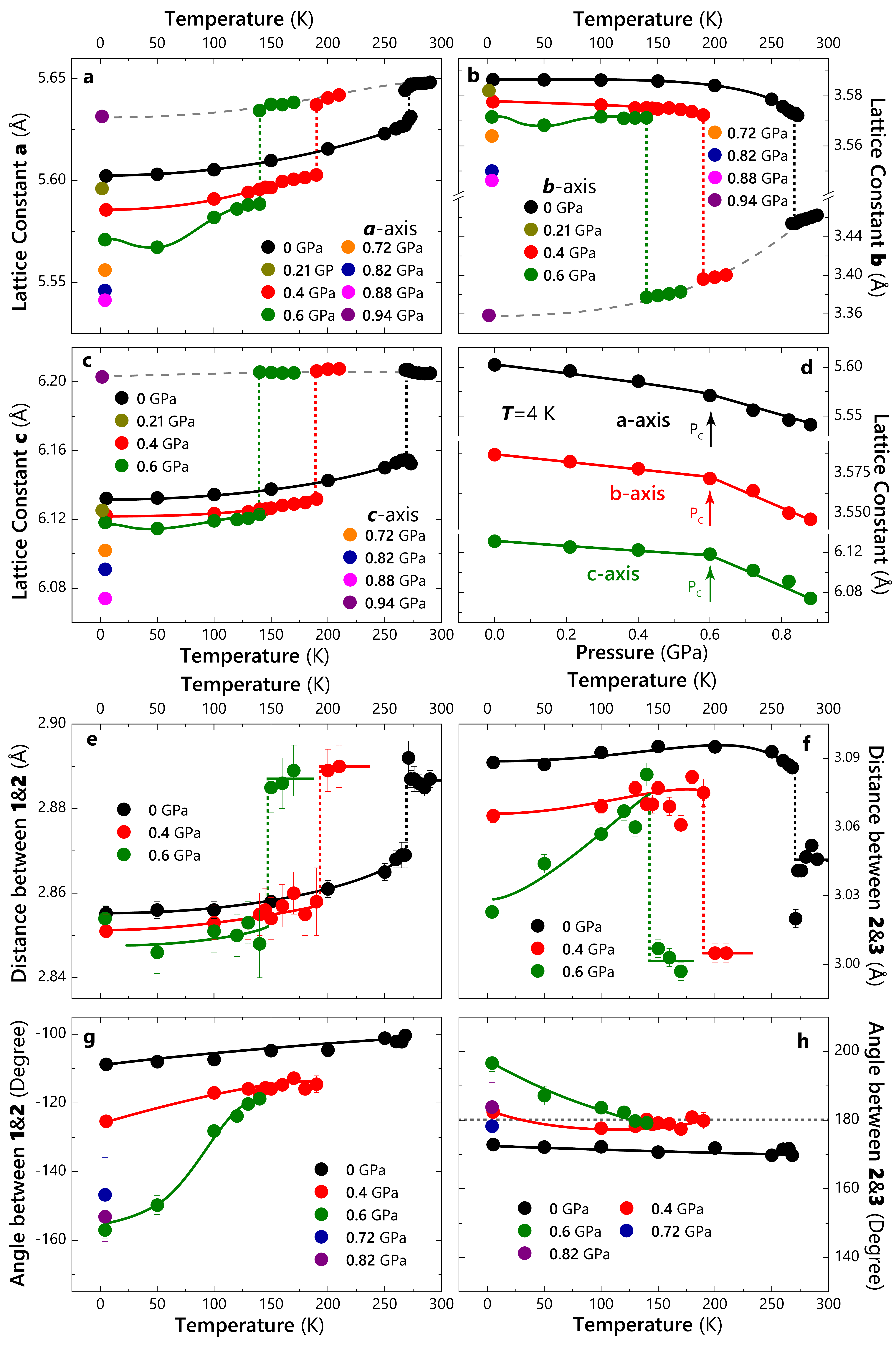}
\caption{\textbf{Temperature and pressure dependence of lattice constants, bond distances, and angles between Cr moments obtained from the refinement analysis.} \textbf{a-c}, Lattice parameter as a function of temperature and pressure. \textbf{d}, Pressure dependence of lattice parameter in the magnetically ordered state. A clear anomaly is observed near the critical pressure $P_c=0.6$ GPa. \textbf{e}, Bond distance between $S_1$ and $S_2$ (nearest-neighbor).  \textbf{f}, Bond distance between $S_2$ and $S_3$ (nearest-neighbor). \textbf{g}, Angle ($\beta_{12}$) between spin $S_1$ and $S_2$. \textbf{h}, Angle ($\beta_{23}$) between spin $S_2$ and $S_3$. By symmetry, we have $\beta_{12}$=$\beta_{34}$, $\beta_{23}$=$\beta_{41^\prime}$ (Figs. 1a and 1b). The error bars indicate one standard deviation.
}
\end{figure}

\clearpage

\begin{table*}
\center
\caption{\label{tab:table1}%
Refined magnetic and structural parameters of CrAs under ambient and applied pressure in the aluminum pressure cell. Space group: Pnma. Atomic positions: Cr: 4c (x, 1/4,  z); As: 4c(x, 1/4, z).
}
\begin{ruledtabular}
\begin{tabular}{clcccccc}
\multirow{2}{1cm}{\textrm{Atom}}& \multirow{2}{1cm}{\textrm{}}& \textrm{$P=0\ GPa$}& \textrm{$P=0\ GPa$}& \textrm{$P=0.4\ GPa$}& \textrm{$P=0.4\ GPa$}& \textrm{$P=0.6\ GPa$}& \textrm{$P=0.6\ GPa$}\\
&& \textrm{$T=4\ K$}& \textrm{$T=290\ K$}& \textrm{$T=4\ K$}& \textrm{$T=210\ K$}& \textrm{$T=50\ K$}& \textrm{$T=170\ K$}\\
\hline
& \textrm{$a({\AA})$} & 5.60499(5) & 5.65101(8) & 5.5882(1) & 5.6448(1) & 5.5700(1) & 5.6411(2)\\
& \textrm{$b({\AA})$} & 3.58827(3) & 3.46398(5) & 3.57937(7) & 3.40203(8) & 3.57003(9) & 3.38452(9)\\
& \textrm{$c({\AA})$} & 6.13519(5) & 6.20804(8) & 6.1253(1) & 6.2106(1) & 6.1176(1) & 6.2082(1)\\
& \textrm{$k$} & 0.35643(7) & - & 0.3171(2) & - & 0.2080(4) & - \\
\textrm{Cr} & \textrm{$x$} & 0.0070(2) & 0.0060(3) & 0.0093(5) & 0.0091(5) & 0.0086(7) & 0.0081(7)\\
& \textrm{$z$} & 0.2049(2) & 0.2019(3) & 0.2029(4) & 0.1994(5) & 0.2015(5) & 0.1992(6)\\
& \textrm{$B({\AA^2})$} & 0.15(2) & 0.74(5) & 0.14(5) & 0.32(6) & 0.08(8) & 0.27(8)\\
& \textrm{$m(\mu_B)$} & 1.724(9) & - & 1.71(2) & - & 1.41(4) & - \\
\textrm{As} & \textrm{$x$} & 0.2045(1) & 0.2016(2) & 0.2026(2) & 0.1985(3) & 0.1990(3) & 0.1979(4)\\
& \textrm{$z$} & 0.5836(1) & 0.5766(2) & 0.5837(3) & 0.5738(3) & 0.5822(4) & 0.5726(4)\\
& \textrm{$B({\AA^2})$} & 0.03(1) & 0.58(3) & 0.44(3) & 0.24(2) & 0.37(6) & 0.17(7)\\
\hline
& \textrm{$Rp(\%)$} & 4.82 & 6.63 & 4.73 & 4.81 & 5.86 & 5.40\\
& \textrm{$wRp(\%)$} & 5.87 & 8.21 & 5.95 & 5.91 & 7.05 & 6.69\\
& \textrm{$\chi^2$} & 1.55 & 1.40 & 2.00 & 1.25 & 1.10 & 0.97\\
\end{tabular}
\end{ruledtabular}
\end{table*}

\begin{table*}
\center
\caption{\label{tab:table1}%
Refined magnetic and structural parameters of CrAs under various pressures in the iron pressure cell.
}
\begin{ruledtabular}
\begin{tabular}{cccccccccccccc}
\textrm{$P$} & \textrm{$T$} & \textrm{$a$} & \textrm{$b$} & \textrm{$c$} & \textrm{$m-Cr$} & \textrm{$k-vector$} & \textrm{$Rp$} & \textrm{$wRp$} & \textrm{$\chi^2$}\\
\textrm{$GPa$} & \textrm{$K$} & \textrm{${\AA}$} & \textrm{${\AA}$} & \textrm{${\AA}$} & \textrm{$\mu_B$} & & \textrm{$\%$} & \textrm{$\%$} & \\
\hline
\textrm{$0.72$} & \textrm{$4$} & 5.556(5) & 3.564(1) & 6.102(2) & 1.25(10) & 0.1752(8) & 2.27 & 2.98 & 2.69\\
\textrm{$0.82$} & \textrm{$4$} & 5.546(3) & 3.550(1) & 6.091(2) & 1.24(3) & 0.1706(8) & 5.63 & 7.74 & 2.44\\
\textrm{$0.88$} & \textrm{$4$} & 5.541(2) & 3.546(2) & 6.074(8) & 1.0(2) & 0.144(6) & 7.28 & 9.85 & 3.73\\
\textrm{$0.94$} & \textrm{$1.5$} & 5.6316(6) & 3.3585(4) & 6.2029(6) & 0 & - & 2.89 & 3.60 & 3.70\\
\end{tabular}
\end{ruledtabular}
\end{table*}

\clearpage
\renewcommand\figurename{\textbf{Figure S}}
\setcounter{figure}{0}

\maketitle{\textbf{Supplementary Information: Structural and Magnetic Phase Diagram of CrAs and its Relationship with Pressure-induced Superconductivity}}

\textbf{I. Exchange model based on nearest neighbor spins and the Dzyaloshinskii-Moriya interactions}

     The rapid decrease of the propagation vector along with the spin reorientation from the $ab$ plane (low pressure phase) to $ac$ plane (high pressure phase) can be qualitatively understood as follows. There are four spins in one unit cell, as labeled in Figs. 1a and 1b. We can consider a minimum magnetic exchange model based on the nearest neighbor spins.  The interaction between the spin $\hat S_i$ and $\hat S_j $ can be generally written as \begin{eqnarray}
\hat H_{ij}=  \vec D_{ij} \cdot (\hat S_i\times \hat S_j)+ J_{ij} \hat S_i\cdot \hat S_j,
\label{h12}
\end{eqnarray}
where  the first term is the Dzyaloshinskii-Moriya interaction; and $\vec D$ is the Dzyaloshinskii-Moriya interaction unit vector.   $\vec D_{ij}$ is proportional to $(\vec R_{As_{ij,1}}+\vec R_{As_{ij,2}}-\vec R_{Cr_i}-\vec R_{Cr_j})\times (\vec R_{Cr_i}-\vec R_{Cr_j})$, where $\vec R_{As_{ij,\alpha}}$ are coordinates of the As atoms that link $\hat S_i$ and $\hat S_j$ spins, as shown in Figs. 1a and 1b.
For $\hat S_1$ and $\hat S_2$,  as the two As atoms and the two Cr spins lack an inversion center, $\vec D_{12}$
is nonvanishing.  However,  for $\hat S_2$ and $\hat S_3$, there is an inversion center so that $\vec D_{23}=0$.  Moreover, by symmetry, we have $H_{12}=H_{34}$.   By plugging in the atom coordinates, we obtain
 $\vec D_{12}=\vec D\sim D_0 (-0.17, -0.5, 0.85)$ in both phases, as the difference of the lattice constants between the two phases is small. Therefore, the minimum model to describe one helix chain along the $c$ axis  is given by
 \begin{eqnarray}
 \hat H& = &\sum_i  \vec D\cdot (\hat S_{i,1}\times \hat S_{i,2}+\hat S_{i,3}\times \hat S_{i,4})+ J_{1} (\hat S_{i,1}\cdot \hat S_{i,2}+\hat S_{i,3}\cdot \hat S_{i,4})\nonumber \\ & + & {J_1}^\prime (\hat S_{i,2}\cdot \hat S_{i,3}+\hat S_{i,1}\cdot \hat S_{i+1,4}).
\label{ht}
\end{eqnarray}
Taking the spin rotation plane in the helical magnetic structure into account, we show that this model can qualitatively explain the two observed magnetic structures and the change of the helical propagation vector between the two phases.
 With ${J_1}^\prime>0$,  it accounts for the antiferromagnetic spin orientation between $S_2$ and $S_3$ observed in both phases.   The angle $\beta_{12}$  between $S_1$ and $S_2$ is given by $ tan (\beta_{12}) =   \vec D\cdot\vec n/J_1$, where $\vec n$ defines the helix spin rotation plane, namely the hard spin axis. The helical propagation vector $\vec k =(0,0, \frac{2|\beta_{12}+\beta_{23}|}{2\pi})$, where $\beta_{23}$ is the angle between $S_2$ and $S_3$.
 In the low pressure phase, $\vec n=(0,0,1)$, and in the high pressure phase, $\vec n =(0,1,0)$. As the component of $\vec D$ along the $c$ axis is larger than the one along the $b$ axis, the helix wavevector decreases if the helix rotation plane changes from the $ab$ plane in the low pressure phase to the $ac$ plane in the high pressure phase, since angle $\beta_{23}$ hardly changes with pressure and is always close to $180^\circ$ in both phases.

\textbf{II. Magnetic susceptibility measurements of polycrystalline CrAs}
\begin{figure}[h]
\includegraphics[scale=0.5]{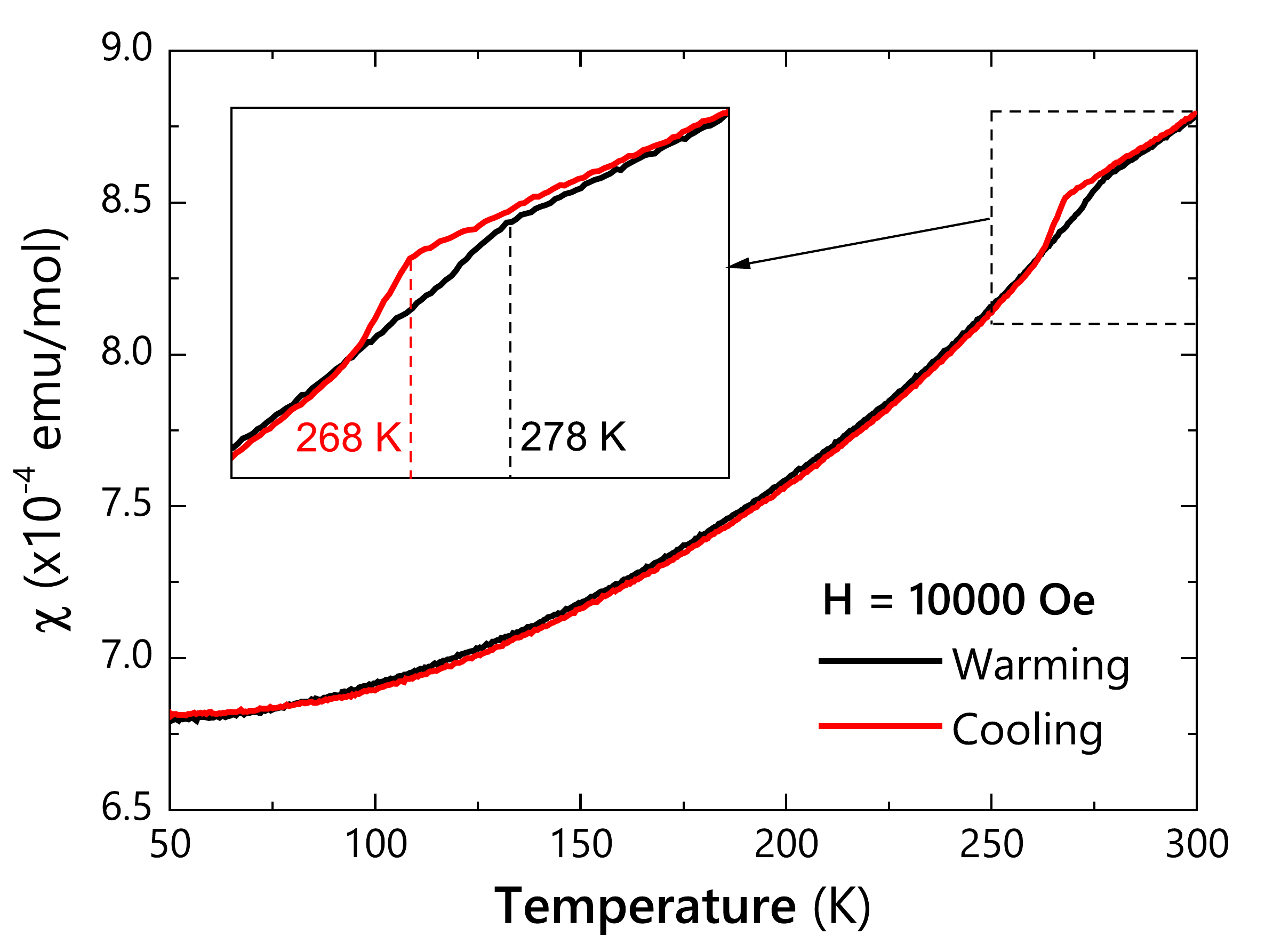}
\caption{\textbf{ Temperature dependence of the magnetic susceptibility of polycrystalline CrAs.} Magnetization measurements were performed in a Quantum Design superconducting quantum interference device (SQUID) magnetometer. The temperature dependence of the magnetic susceptibility displays a clear anomaly near $270$ K. The inset shows the the thermal hysteresis near 270 K.
}
\end{figure}

\end{document}